# Controlling an altermagnetic spin density wave in the kagome magnet CsCr$_3$Sb$_5$

Zihao Huang[1,#], Chenchao Xu[2,3,#], Yande Que[1,#], Yi Liu[3,4,#], Ranjith Shivajirao[1], Zheng Jue Tong[1], Amit Kumar[1], Chao Cao[3,*], Guang-Han Cao[3,*], Hong-Jun Gao[5], Bent Weber[1,*]

[1] School of Physical and Mathematical Sciences, Nanyang Technological University, Singapore 637371, Singapore

[2] School of Physics, Hangzhou Normal University, Hangzhou 310036, P. R. China.

[3] School of Physics, Zhejiang University, Hangzhou 310027, PR China.

[4] Department of Applied Physics, Key Laboratory of Quantum Precision Measurement of Zhejiang Province, Zhejiang University of Technology, Hangzhou, PR China.

[5] Beijing National Center for Condensed Matter Physics and Institute of Physics, Chinese Academy of Sciences, Beijing 100190, PR China

[#]These authors contributed equally to this work

[*]Correspondence author, Email address: *b.weber@ntu.edu.sg*; *ghcao@zju.edu.cn*; *ccao@zju.edu.cn*

The interplay of charge and spin orders lies at the heart of correlated electron physics and plays a critical role in the emergence of unconventional quantum phases. Kagome magnets provide a particularly promising platform to investigate these phenomena, owing to their geometrically frustrated lattice structure. However, resolving spin and charge orders microscopically and establishing ways to control them remain fundamental challenges. Here, we demonstrate magnetic-field control of an altermagnetic spin density wave order intertwined with charge density wave order in kagome magnet $CsCr_3Sb_5$. Scanning tunneling microscopy down to deep cryogenic temperature of 50 mK reveals two previously unreported charge density wave orders in the Sb surface. Density functional theory confirms that one of them, a $4a_0 \times \sqrt{3}a_0$ charge order, is coupled to a spin density wave with an altermagnetic ground state. The charge density waves can be tuned in both amplitude and phase by an external magnetic field, reflected in domain switching and stripe sliding of the charge density wave. Our findings deepen the understanding of symmetry-breaking in kagome systems, providing a tunable platform to explore the interplay of electronic correlation with emergent quantum magnetism.

Emergent charge and spin order are central themes in condensed matter physics as they reflect the behavior of correlated electrons and often serve as precursors or competitors to unconventional superconductivity[1-6]. The interplay between these collective orders has been linked to phenomena such as pseudogaps[7,8], nematicity[9,10], and quantum criticality[11,12]. Probing how charge density waves (CDWs) and spin density waves (SDWs) emerge, interact, or compete is critical to build an understanding of correlated quantum phases, and to develop ways of tuning and controlling these exotic states.

Kagome materials offer a compelling platform for this exploration. Their geometrically frustrated lattice supports flat bands, enhances correlation effects, and promotes complex symmetry-breaking orders[13-15]. Recently, the family of kagome materials $AV_3Sb_5$ (A=K, Rb, Cs) has attracted considerable attention for its diversity of emergent phases, including unconventional CDWs[16,17], electronic nematicity[18], superconductivity[19,20], and pair density waves[21], though notably without spin magnetism. By contrast, the isostructural compound $CsCr_3Sb_5$ has recently drawn interest for introducing spin magnetism and exhibiting a rich phase diagram of correlated states[22-31]. Angle-resolved photoemission spectroscopy (ARPES) and first-principles density functional theory (DFT) calculations reveal nearly flat bands near the Fermi level, suggesting the presence of strong electronic correlation[23,24,26,27,30,31]. At ambient pressure, it exhibits an antiferromagnetic transition at ~55 K, accompanied by a unidirectional density wave order[22]. Optical spectroscopy suggests the density wave may coexist with a nematic electronic state[29], while applied pressure suppresses the density wave and induces a superconducting phase near a putative quantum critical point[22].

The density wave order stands out as both foundational and enigmatic. On the one hand, it appears to compete with the superconductivity in the pressure phase diagram[22], yet its magnetic fluctuations may provide the pairing glue for unconventional superconductivity[23,25]. On the other hand, the precise nature of the density wave — whether it is long-range or short-range, or driven by electron-phonon or electron-electron interactions — remains under active debate[22,28]. Moreover, recent first-principles calculations identify an altermagnetic spin density wave ground state in $CsCr_3Sb_5$[25]—a novel type of magnetic order characterized by compensated antiferromagnetic spin polarization and momentum-dependent band splitting[32-36]. Thus, investigating the microscopic nature of the density wave order and its connection to magnetism in $CsCr_3Sb_5$ is crucial and urgent.

In this work, using ultralow-temperature (T~50 mK) scanning tunneling microscopy/spectroscopy (STM/STS) under external magnetic fields, we report two previously unreported CDW orders intertwined with tunable altermagnetic SDW order in CsCr$_3$Sb$_5$. Pronounced peaks in the tunneling spectra are consistent with flat bands near the Fermi level, which promote strong electronic correlations. DFT calculations show that one of these CDW orders, a $4a_0 \times \sqrt{3}a_0$ modulation, is accompanied by a SDW with an altermagnetic ground state. Consistent with this notion, we observe that both the phase and the amplitude of these CDWs can be tuned by an applied magnetic field, accompanied by reorganization of the CDW domains. Within well-defined domains, we further observe a sliding CDW behavior, where the phase evolves continuously with magnetic field as the rate of $0.12\pi/T$. These results establish CsCr$_3$Sb$_5$ as a promising altermagnetic SDW candidate and kagome platform to explore magnetic-field-tunable correlated states.

**Surface atomic structure and flat band features**

CsCr$_3$Sb$_5$ shares the same crystallographic structure as its nonmagnetic counterpart CsV$_3$Sb$_5$[37], featuring a layered stacking sequence: Cs-Sb2-CrSb1-Sb2-Cs (Fig. 1**a**), with space group No.191, P 6/*mmm*. Due to the weak chemical bonding between Cs and Sb2 layers, cleaving typically exposes either Cs-terminated or Sb2-terminated surfaces. STM topographs reveal that Cs-terminated surfaces display a triangular lattice, and a domain region with $\sqrt{3} \times \sqrt{3}$ periodic Cs-reconstruction (Fig. S1), likely arising from surface atom rearrangement. The Sb2-terminated surfaces exhibit a honeycomb lattice (Figs. 1**b**, **c**), consistent with the atomic structure expected for this layer and an in-plane lattice constant of $a_0 \sim 5.5$ Å[22]. Due to the mobility of Cs atoms[38, 39] on the cleaved surface, most measurements were performed on the more stable Sb2-terminated surfaces (see method for the surface preparation).

The kagome lattice is known to host flat bands in its electronic structure due to geometric configuration and interference effects. In CsCr$_3$Sb$_5$, DFT calculations have predicted the presence of several nearly flat bands in close proximity to the Fermi level $E_F$[22, 23, 26, 27]. Our DFT calculations of the surface electronic structure reveals multiple flat bands (Fig. 1**d**), as visible from the pronounced peaks in the corresponding density of states (Fig. 1**e** blue). Differential conductance (d*I*/d*V*) spectra acquired on the Sb2-terminated surface (Fig. 1**e** red) exhibit prominent peaks at around –251 mV, –60 mV, 315 mV, and 515 mV (indicate by arrows), in good agreement with the energy of these flat bands.

Notably, the spectral weight near the $E_F$ is significantly suppressed, expected for the opening of a CDW gap, which suppresses the density of states around $E_F$. Indeed, high-resolution spectra near $E_F$ (Fig. 1**g**), measured at 50 mK ($T_{eff}$ ~ 150 mK), reveal a gap of approximately 9 meV, consistent with a CDW-induced spectral suppression. A unidirectional contrast pattern is already visible in the high-bias STM topographic image (Fig. 1**b**), indicative of the underlying CDW order. As the bias is decreased, two distinct unidirectional stripe modulations with different periodicities emerge (Fig. 1**f**). Down to the lowest accessible measurement temperature, we do not observe any features consistent with a superconducting state (inset of Fig. 1**g**).

**Two competing CDW orders**

To confirm the presence of CDW orders, we acquired low-bias topographic images across multiple samples and consistently observed two distinct stripe-like patterns (Figs. 2**a**, 2**d**), of which both break translation and rotational symmetry of the lattice. The first CDW order exhibits stripe modulation with $4a_0 \times \sqrt{3}a_0$ periodicity (Fig. 2**a**). The corresponding Fourier transform (FT) reveals wavevectors $\boldsymbol{q_{4a_0}}$ and $\boldsymbol{q_{\sqrt{3}a_0}}$ (Fig. 2**b**; See the schematic illustration in Fig. S2), both of which are non-dispersive in energy, as demonstrated by bias-independent FT line-cuts along the $\Gamma - K$ and $\Gamma - M$ directions (Fig. 2**c**). This confirms the energy-independent nature of the modulations, excluding quasiparticle interference as their origin. The additional, weaker scattering vectors can be attributed to higher-order harmonics of the primary wavevectors, as well as wave vector mixing[40, 41] components involving the primary CDW vectors and the Bragg peaks. Specifically, these mixed components take the form $\vec{q}_{mix} = m \cdot \vec{Q}_{Bragg} + n \cdot \vec{q}_{CDW}$ ($m, n \in \mathbf{Z}$).

A second CDW stripe pattern displays a longer periodicity with $8a_0 \times \sqrt{3}a_0$ (Fig. 2**d**). Similar analysis reveals corresponding wavevectors $\boldsymbol{q_{8a_0}}$ and $\boldsymbol{q_{\sqrt{3}a_0}}$ that are also non-dispersive (Figs. 2**e–f**). Notably, the coexistence of these two CDW orders in adjacent domains within the same field of view (Fig. 1**e**) suggests that they represent energetically competing phases, possibly stabilized by subtle variations in surface conditions or local defects. In fact, throughout most of the field of view, these two CDW stripe patterns commonly coexist within small domains, as exemplified in Fig. 1**f**. The presence of multiple domains with different stripe orientations (Fig. S3) further implies the breaking of the six-fold symmetry of the lattice and suggests that the CDW state selects a specific wavevector direction.

Notably, these modulations differ from a $4a_0 \times 1a_0$ CDW order previously reported by X-ray diffraction (XRD) measurements[22, 28] of the CsCr$_3$Sb$_5$ bulk. This underscores the importance of surface-sensitive, atomic-resolution microscopy in real space, which enables the attribution of the observed charge orders to states emerging predominantly at or near the surface. The possibility of surface reconstruction is excluded, as high-bias STM images (Fig. 1**b**) show an undistorted honeycomb structure. To ultimately confirm the CDW nature, we performed d$I$/d$V$ mapping at positive and negative biases. The electron and hole components of the order parameter exhibit a phase shift, as expected for CDW systems, an reflected in a contrast inversion[42, 43]. For both the $4a_0 \times \sqrt{3}a_0$ (Fig. S4**a** and Fig. 2**g**) and $8a_0 \times \sqrt{3}a_0$ (Fig. S4**b** and Fig. 2**h**) CDWs, we observe clear inversion of contrast upon bias reversal, consistent with the particle-hole symmetry breaking.

Although the reported bulk $4a_0 \times 1a_0$ CDW is absent at the surface, we attempt to probe its signature by reducing the tip-sample distance to access tunneling electrons from deeper layers. At large tip-sample distances (Figs. S5**a** and **b**), no scattering vector is observed near the expected $1/4|Q_{Brag}|$ position, within the noise background. However, at smaller tip-sample distances (Figs. S5**c-e**), a very faint scattering vector emerges at the expected $1/4|Q_{Brag}|$ position, accompanied by its high order components due to mix with $q_{8a_0}$. This comparison across different tip-sample distances further confirms that the $4a_0 \times 1a_0$ CDW order is present within bulk, but is not favored at the surface. A similar surface–bulk difference has been reported in CsV$_3$Sb$_5$, where the $4a_0$ charge order[16, 17] has been considered surface-derived[44]. In addition, the absence of the surface Cs layer, which normally donates electrons, may further contribute to the difference between surface and bulk charge orders[45 – 47].

**Altermagnetic SDW ground state and its interplay with CDW order**

To further confirm the origin of the observed CDW states, we performed DFT calculations (see Methods) on the CsCr$_3$Sb$_5$ surface. Towards this end, we consider a Cr$_3$Sb$_5$ monolayer, simulating the Sb2-terminated surface after Cs removal, consistent with the experiment. A total of 760 candidate magnetic textures-constrained by symmetry of the observed $4a_0 \times \sqrt{3}a_0$ CDW configuration-were examined and their total energies compared (Fig. S6**a**). The SDW states that were reported as the lowest-energy configurations of the CsCr$_3$Sb$_5$ bulk[25] are also included.

Among these, a $4a_0 \times \sqrt{3}a_0$ SDW at the surface was identified as the ground state with the lowest energy, whose spin texture is illustrated in Fig. 3**a**. Here, the alternating spin orientations on the Cr kagome lattice form a compensated but time-reversal-symmetry-breaking altermagnetic order configuration. The two spin-sublattices are not connected spatially by either simple translation or inversion. Instead, the sublattices are connected by a $\{M_{yz}|(0,1/2,0)\}$ symmetry operation, in which Cr sites with opposite spin but same label index are connected through a combination of mirror symmetry about the *yz*-plane ($M_{yz}$) and a half translation along the *y*-axis. The corresponding spin-space group is $P^1m^{-1}c^{-1}2_1\ ^{\infty}m1^{48,\,49}$. The calculated magnetic moment distribution within the Cr kagome layer is shown in Fig. 3**b** (see also in Supplementary Table 1).

Based on this lowest-energy SDW configuration, we have further calculated the corresponding charge density modulation (see Methods) to compare with our STM topographic images (Fig.3**c**). The calculated spatial charge distribution (lower panel of Fig.3**c**) reproduces the observed $4a_0 \times \sqrt{3}a_0$ periodicity and real-space pattern (upper panel of Fig.3**c**) extremely well, demonstrating a direct connection between the CDW and the underlying SDW. Importantly, the glide-mirror symmetry of the SDW order is also inherited by the CDW, where the intracell modulation pattern can likewise be generated through the combined mirror and translation operations (Fig. 3**c**). We have also tested several other SDW configurations with total energies slightly above that of the ground state (Figs. S6**b-h**). However, none of their calculated charge density patterns match the experimental topography, reinforcing our identification of the altermagnetic SDW ground state. The altermagnetic character of the SDW is further reflected in a clear momentum dependent spin-splitting as confirmed by the calculated Fermi surface (Fig. 3**d**), unambiguously confirming the key features of altermagnetic order.

**Magnetic field tunability and sliding CDW order**

To further demonstrate how the altermagnetic SDW and CDW orders are intertwined, we have performed STM topographic measurements recorded under application of an out-of-plane magnetic field ($B_z$). Scans within the same field of view (Fig. 4**a**) reveal pronounced changes in the CDW patterns, including splitting (Fig. S7), shifting (Fig. S8), and interconversion between the $4a_0$ and $8a_0$ modulations (Fig. S9). By applying a Fourier filter, leaving only the $\boldsymbol{q}_{4a_0}$ and $\boldsymbol{q}_{8a_0}$ scattering vectors and performing an inverse Fourier transform, we can isolate the stripe components (Fig. 4**b**), which clearly illustrate their field-dependent evolution. A temporal origin can be excluded, as the CDW stripes remain stable over 12 hours intervals (Figs. S10**c**, **d**) in the absence of a magnetic field.

To accurately track changes of the CDW/SDW domain structure, we carried out two-dimensional (2D) lock-in analysis (see Methods) on the $4a_0$ stripes using the magnetic-field-dependent topographic images. This method yields the spatial distributions of the CDW amplitude $A(r)$ (Fig. 4**c**) and phase $\phi(r)$ (Fig. 4**d**), both of which are seen to clearly evolve with applied magnetic field.

Finally, to track changes within a single domain, we followed domain #1 under low magnetic field of 0–1.5 T, at which the domain remains stable without structural change. We observe that the stripes evolve continuously without breaking into smaller domains (Fig. 5**a**) and show a progressive lateral shift with respect to their 0 T positions. The corresponding filtered images (Fig. 5**b**) and line profiles (Fig. 5**c**) confirm this shift. To further quantify this effect, we fit the line profiles with $\cos(q_{4a_0} \cdot r + \varphi)$ to extract the stripe phase $\varphi$. Data across four different domains that display continuous evolution within accessible magnetic field ranges (Fig. S11) are summarized in Fig. 5**d**, showing a monotonic evolution and suggesting a magnetic field induced sliding of the CDW/SDW at a rate of $0.12\pi/T$. We note that similar magnetic-field-driven evolution is also observed for the $\sqrt{3}a_0$ modulation (Figs. S12 and S13).

## Discussion

The magnetic field–induced evolution of CDW patterns observed is highly unconventional. In typical systems, CDW states arise from charge modulation due to Fermi surface nesting or electron–phonon coupling and do not possess intrinsic magnetic components. As such, CDWs are generally expected to be insensitive to external magnetic fields. In contrast, our STM measurements reveal significant changes in both the amplitude and phase of the CDW under modest magnetic fields, strongly suggesting that the CDW in CsCr$_3$Sb$_5$ is intertwined with the underlying altermagnetic SDW order.

This idea is supported by recent work on other kagome systems, such as GdTi$_3$Bi$_4$[50], in which magnetic field–induced distortions of CDW patterns were attributed to intertwined CDW–SDW states. Transport experiments on CsCr$_3$Sb$_5$ have reported an antiferromagnetic transitions associated with the CDW near 55 K, implying that spin order is an integral part of the low-temperature ground state. Our first-principles calculations support this perspective by showing that the energetically favored SDW ground state at the surface indeed exhibits altermagnetic order with a $4a_0 \times \sqrt{3}a_0$ periodicity.

In many systems, local magnetic moments can tilt, flip, or reorient under just a few tesla of applied field[51-53]. This field sensitivity could be expected to modify the SDW texture, therefore alter the CDW order that is intertwined with the SDW. Our real-space observations of field-tunable stripe displacement and amplitude variation are consistent with such a scenario. It should also be noted that real material surfaces are often complicated by domain walls, defects, and strain, all of which can influence the SDW texture and its field response. These factors may contribute to the continuous phase evolution that deviates from perfect commensurability (Fig. 5**d**).

Another possible mechanism involves the intrinsic electronic structure of the SDW state. Our DFT calculations support a commensurate SDW ground state giving rise to altermagnetic orders. In this interpretation, an applied magnetic field introduces additional Zeeman shift on the spin-polarized Fermi surfaces. This field-induced Fermi surface reconstruction may destabilize or reshape the SDW texture, which in turn modifies the charge distribution and leads to the observed evolution of the CDW. Finally, the interplay between distinct bulk-surface CDW behaviors and coupled CDW/SDW orders positions kagome monolayers as a promising platform for exploring novel device functionalities.

In conclusion, we report two previously unobserved CDW modulations in $CsCr_3Sb_5$ intertwined with an altermagnetic SDW order whose amplitude and phase are tunable by an external magnetic field. These findings uncover a unique example of field-controllable charge order arising from its coupling to an altermagnetic spin texture. These intertwined orders not only enrich the landscape of symmetry-breaking phenomena in kagome materials, but also demonstrate a new platform to study altermagnetism and how magnetic topology and electronic correlations give rise to emergent tunable states. Our work thus offers new insights into magnetically coupled electronic orders, with broader implications for understanding coupled orders in other frustrated or correlated systems, including high-temperature superconductors and topological magnets.

## Materials and Methods

**Single crystal growth of the CsCr$_3$Sb$_5$ sample**[22]. Single crystals of CsCr$_3$Sb$_5$ were synthesized using a self-flux method with high-purity starting materials: Cs (Alfa, 99.999%), Cr (Alfa, 99.99%), and Sb (Aladdin, 99.999%). Eutectic composition from the CsSb–CsSb$_2$ quasi-binary system was used as the flux. The elements were mixed in a molar ratio of 9:2:18 (Cs:Cr:Sb) and loaded into an alumina crucible, which was then sealed in a Ta tube via arc welding under an argon atmosphere. To prevent oxidation, the Ta tube was further sealed in an evacuated silica ampoule. The assembly was gradually heated to 900 °C in a furnace and held at that temperature for 18 hours, followed by slow cooling to 600 °C at a rate of 2 °C per hour. After the growth, the residual flux was removed by immersing the melt in water-free ethanol at room temperature for 48 hours, yielding thin, hexagonally shaped crystalline flakes.

**Scanning tunneling microscopy/spectroscopy.** The samples used in the STM/S experiments were cleaved at low temperature (80 K) and immediately transferred to an STM chamber. Experiments were performed in an ultrahigh vacuum (1×10$^{-10}$ mbar) ultra-low temperature STM system equipped with 11 T magnetic field. All the scanning parameters (setpoint voltage and current) of the STM topographic images are listed in the figure captions. The base temperature is 50 mK in the low-temperature STS. Unless otherwise noted, the d$I$/d$V$ spectra were acquired by a standard lock-in amplifier at a modulation frequency of 726 Hz. Pt-Ir tips were used and calibrated on a clean Au(111) surface prepared by repeated cycles of sputtering with argon ions and annealing at 500 °C. The Cs adatoms at as-cleaved Sb surface were moved away by the STM tip to form a large-scale and clean Sb surface[20,54].

**DFT calculation.** The first-principles calculations were performed based on density functional theory (DFT) with VIENNA AB INITIO SIMULATION PACKAGE (VASP)[55,56]. The exchange-correlations function was taken within the generalized gradient approximation (GGA) in the parameterization of Perdew, Burke and Ernzerhof[57]. The surface was modelled for the Sb2-terminated Cr$_3$Sb$_5$ monolayer, with a 40 Å vacuum layer along z-direction introduced to minimize interlayer interactions. We performed high-throughput search[25] for all possible collinear AFM magnetic configuration within $4a_0 \times \sqrt{3}a_0$ monolayer supercell. Throughout the calculation, the PBEsol approximation was used[58], and spin-orbit coupling (SOC) was not included. The atomic positions and in-plane lattice constants of the Cr$_3$Sb$_5$ monolayer were relaxed using a Γ-centered 2×5×1 k-mesh and a plane-wave cutoff energy of 450 eV. With the optimized lattice parameters, self-consistent calculations and band structure were performed for the $4a_0 \times \sqrt{3}a_0$ SDW configuration with lowest energy. The corresponding unfolded band structure in the Brillouin of the kagome unit cell was obtained after taking averaged from three inequivalent k-paths $\Gamma - K - M - \Gamma$ of the orthorhombic $4a_0 \times \sqrt{3}a_0$ magnetic cell. A tight-binding (TB) Hamiltonian with Cr-3d and Sb-5p was fitted from DFT band structure with Wannier function method[59], with which we obtained the 2D Fermi surfaces.

In order to obtain an improved electronic structure description of the [001] surfaces of $Cr_3Sb_5$ monolayer, a denser k-point mesh with $\Gamma$-centered 8×20×1 was adopted. The experimental STM images can be well reproduced when assuming a tip–sample distance as 2 Å (a vacuum distance above the Sb2-termination of $Cr_3Sb_5$ monolayer) and local density of state (LDOS) integrating from from $E_F$-5 meV to $E_F$ in our calculation.

**2D lock-in technique.** We extract the amplitude $|A_q(r)|$ and phase $\phi_q(r)$ at the scattering vector $q$ by applying a 2D lock-in technique[60–62]. For a real space image $I(r)$, the lock-in image is calculated as:

$$I_q(r) = \int dR I(R) e^{iq \cdot R} e^{-\frac{(r-R)^2}{2\sigma^2}}$$

Then the amplitude and phase are given by:

$$|A_q(r)| = \sqrt{ReI_q(r)^2 + ImI_q(r)^2}$$

$$\phi_q(r) = tan^{-1} \frac{ImI_q(r)}{ReI_q(r)}$$

**Data availability:**

Data measured or analyzed during this study are available from the corresponding author on reasonable request.

# Main Figures

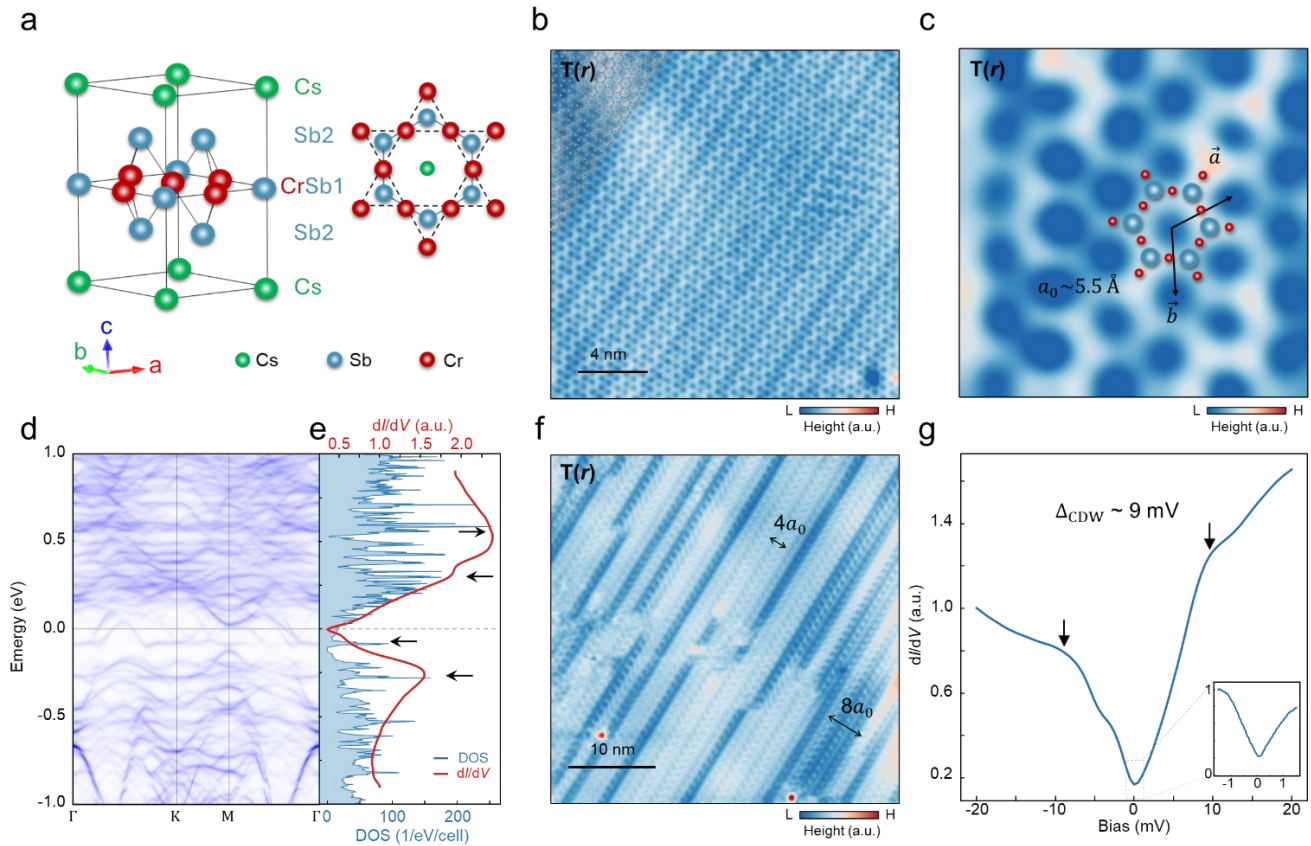

**Fig. 1. Surface atomic structure and flat band features**. **a**, Schematic illustration of crystal structure of kagome metal $CsCr_3Sb_5$. **b**, STM topography of Sb2 surface. Inset shows honeycomb Sb2 lattice. Measurement conditions: $V_b$=-800 mV, $I$=1 nA. **c**, STM topography of Sb2 surface with honeycomb Sb2 lattice. The overlaid Cr atoms lie beneath the Sb2 layer. **d**, Calculated unfolded band structure in the Brillouin of the kagome unit cell of Sb-terminated surface. **e**, Calculated density of states (DOS) and spatial-averaged d$I$/d$V$ spectrum measured on Sb2 surface. Several peaks around Fermi level suggest possible flat bands. Measurement conditions: $V_b$=900 mV, $I$=1 nA, $V_{lock-in}$=10 mV. **f**, STM topographic image of a region with coexistence of two different type of stripe modulation. Measurement conditions: $V_b$=-50 mV, $I$=1 nA. **g**, d$I$/d$V$ spectrum measuring in smaller bias range, showing gap-like feature around Fermi level. Measurement conditions: $V_b$=-50 mV, $I$=2 nA, $V_{lock-in}$=0.5 mV. Inset shows the high-resolution d$I$/d$V$ spectrum. Measurement conditions: $V_b$=5 mV, $I$=200 pA, $V_{lock-in}$=0.03 mV.

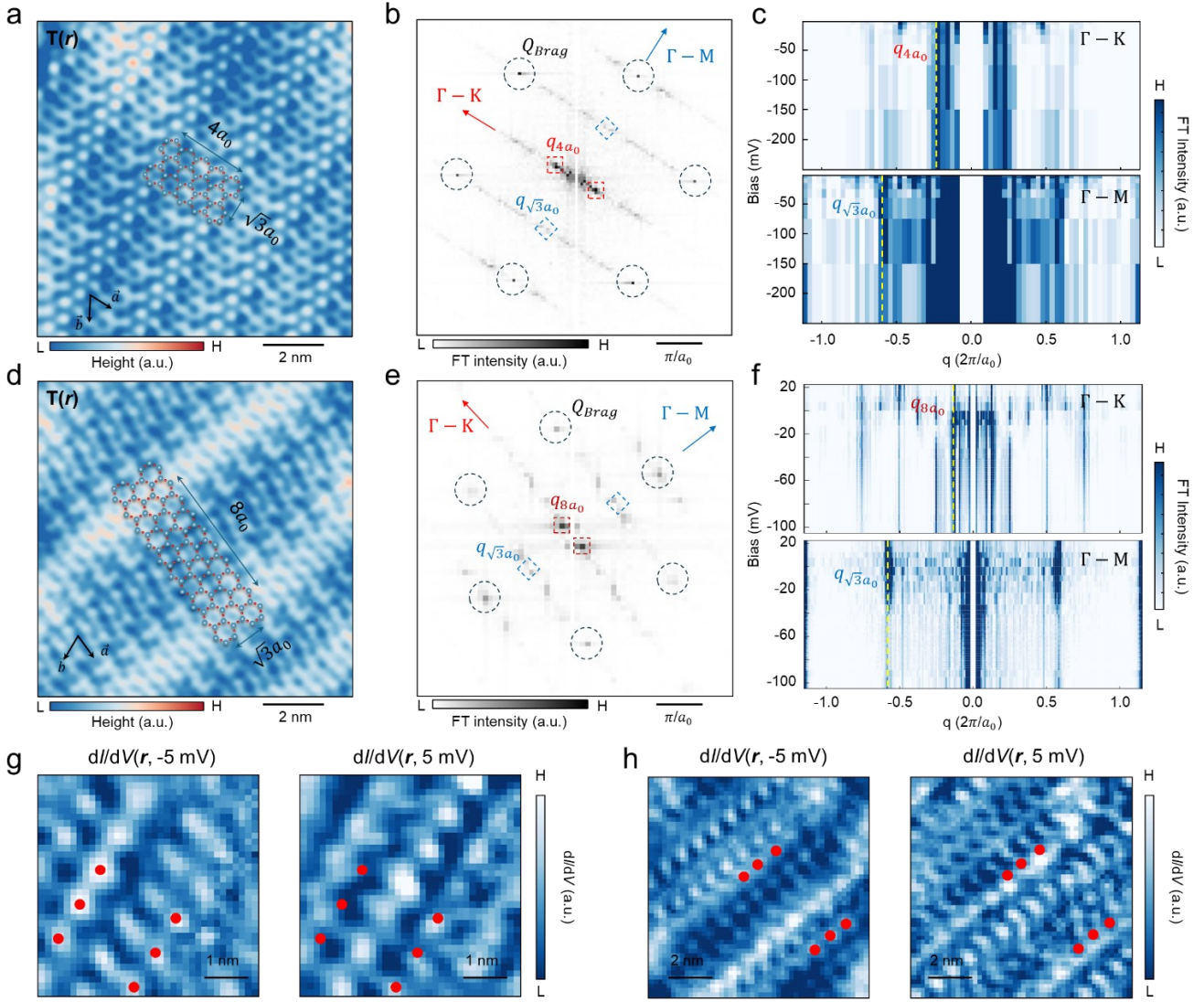

**Fig. 2. Two competing CDW orders**. **a**, STM topographic image of the Sb2 surface with $4a_0 \times \sqrt{3}a_0$ CDW. Sb atoms is overlaid for illustration. Measurement conditions: $V_b$=-50 mV, $I$=1 nA. **b**, FT of (a). The $Q_{Brag}$, $q_{4a_0}$, and $q_{\sqrt{3}a_0}$ are marked by dotted circles, rectangles, and rhombuses, respectively. **c**, Bias-dependent linecuts in the FTs of topography along $\Gamma - K$ (top) and $\Gamma - M$ (bottom) directions, showing the non-dispersive scattering vectors of $q_{4a_0}$, and $q_{\sqrt{3}a_0}$. **d**, STM topographic image of the Sb2 surface with $8a_0 \times \sqrt{3}a_0$ CDW. Sb atoms is overlaid for illustration. Measurement conditions: $V_b$=-30 mV, $I$=1 nA. **e**, FT of (d). The $Q_{Brag}$, $q_{8a_0}$, and $q_{\sqrt{3}a_0}$ are marked by dotted circles, rectangles, and rhombuses, respectively. **f**, Bias-dependent linecuts in the FTs of topography along $\Gamma - K$ (top) and $\Gamma - M$ (bottom) directions, showing the non-dispersive scattering vectors of $q_{8a_0}$, and $q_{\sqrt{3}a_0}$. **g**, d$I$/d$V$ map measured at the region with $4a_0 \times \sqrt{3}a_0$ CDW at -5 mV (left) and +5 mV (right). Measurement conditions: $V_b$=-20 mV, $I$=1 nA, $V_{lock-in}$=0.3 mV. **h**, d$I$/d$V$ map measured at the region with $8a_0 \times \sqrt{3}a_0$ CDW at -5 mV (left) and +5 mV (right). Measurement conditions: $V_b$=90 mV, $I$=1 nA, $V_{lock-in}$=1 mV.

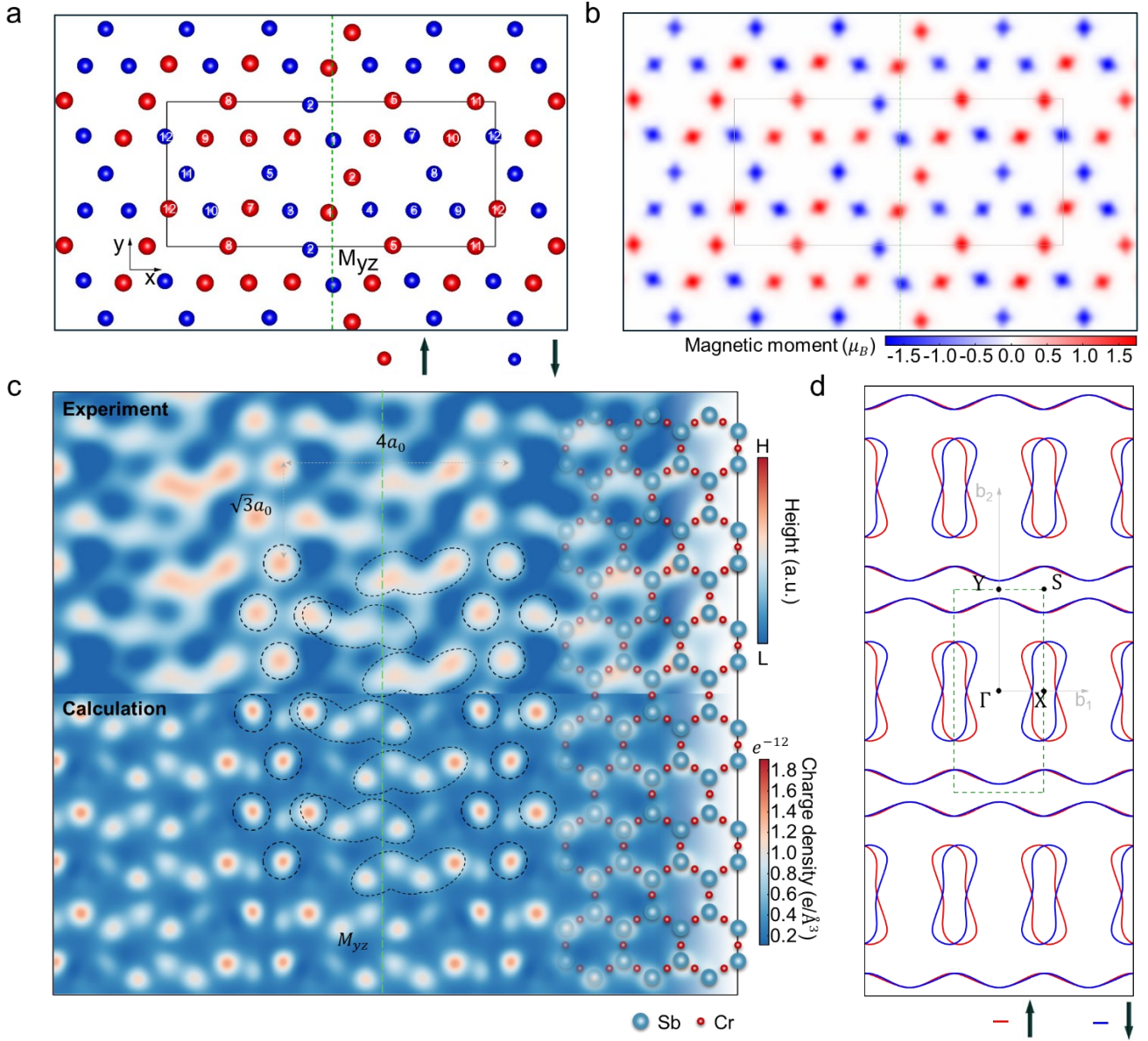

Fig. 3. Altermagnetic SDW ground state and its interplay with CDW order. a, Schematics of real-space spin configuration of the $4a_0 \times \sqrt{3}a_0$ SDW ground state, illustrating an altermagnetic order. Opposite spin orientations are denoted by different atom colors, with symmetry-related Cr sites of opposite spin labeled by the same index. b, Calculated magnetic moment distribution in the Cr kagome layer, consistent with the configuration in (b). c, STM topography of the $4a_0 \times \sqrt{3}a_0$ CDW (top) compared with the calculated charge density modulation (bottom) from the $4a_0 \times \sqrt{3}a_0$ SDW ground state, showing strong agreement in spatial charge patterns. The overlaid Cr atoms are underneath the Sb layer. d, Calculated Fermi surface of the $4a_0 \times \sqrt{3}a_0$ SDW state, exhibiting momentum-dependent spin splitting characteristic of altermagnetism.

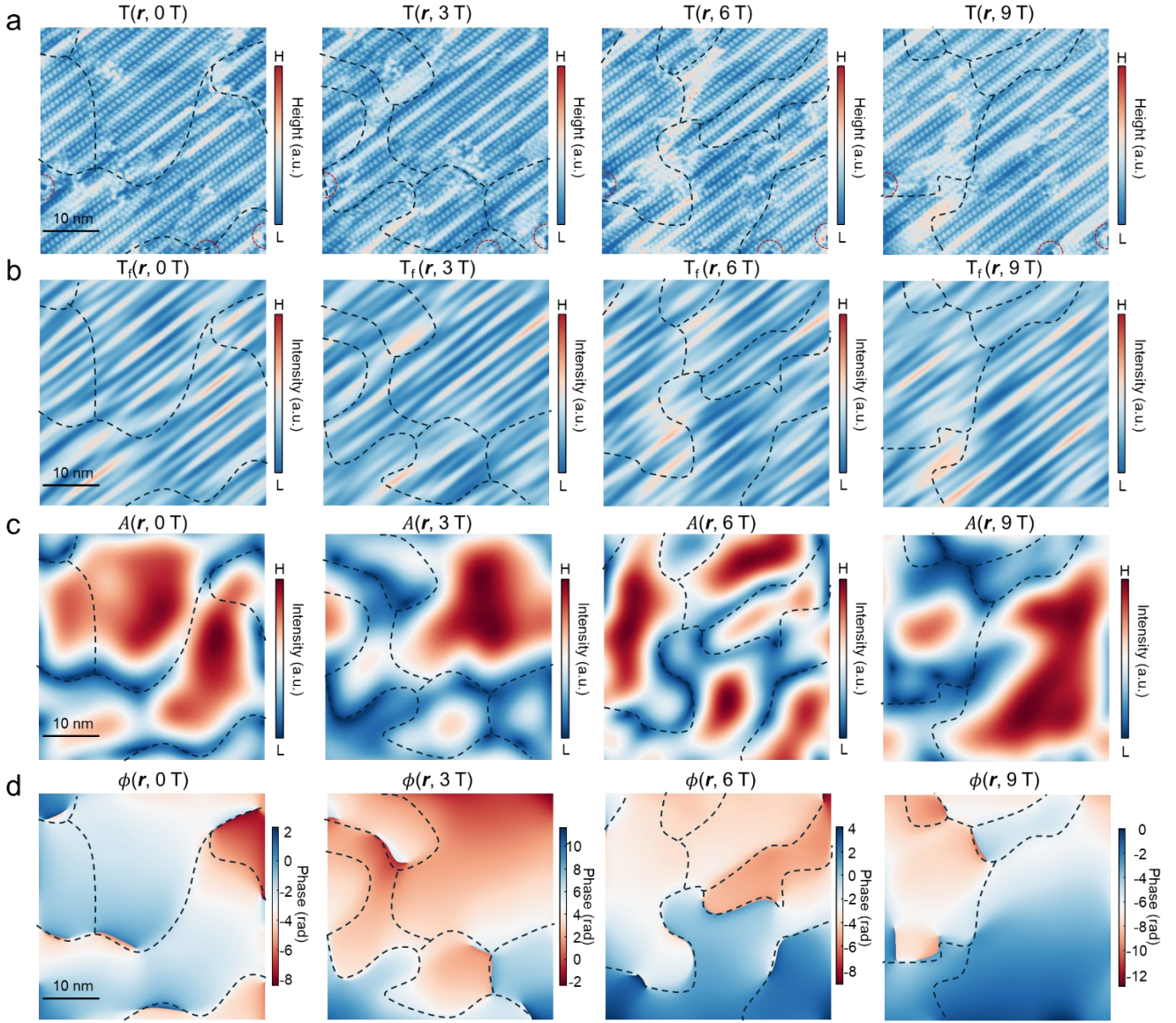

**Fig. 4. Magnetic-field tunability of CDW domains**. **a**, STM topographic images T($r$, $B_z$) of a mixed $4a_0 \times \sqrt{3}a_0$ and $8a_0 \times \sqrt{3}a_0$ CDW region under out-of-plane fields $B_z$ at 0, 3, 6, and 9 T. Measurement conditions: $V_b$=-50 mV, $I$=1 nA. **b**, Filtered topographic image T$_f$ ($r$, $B_z$) obtained by isolating $q_{4a_0}$ and $q_{8a_0}$ and performing inversed FT, from the same regions as in (a). **c**, Spatial maps of $4a_0$ stripe amplitude $A(r, B_z)$, extracted using 2d lock-in technique. **d**, Spatial maps of the unwrapped phase $\phi(r, B_z)$ of $4a_0$ stripes, extracted using 2d lock-in technique. Dotted curves mark stripe domains. The average length of lock-in technique is 3.18 nm.

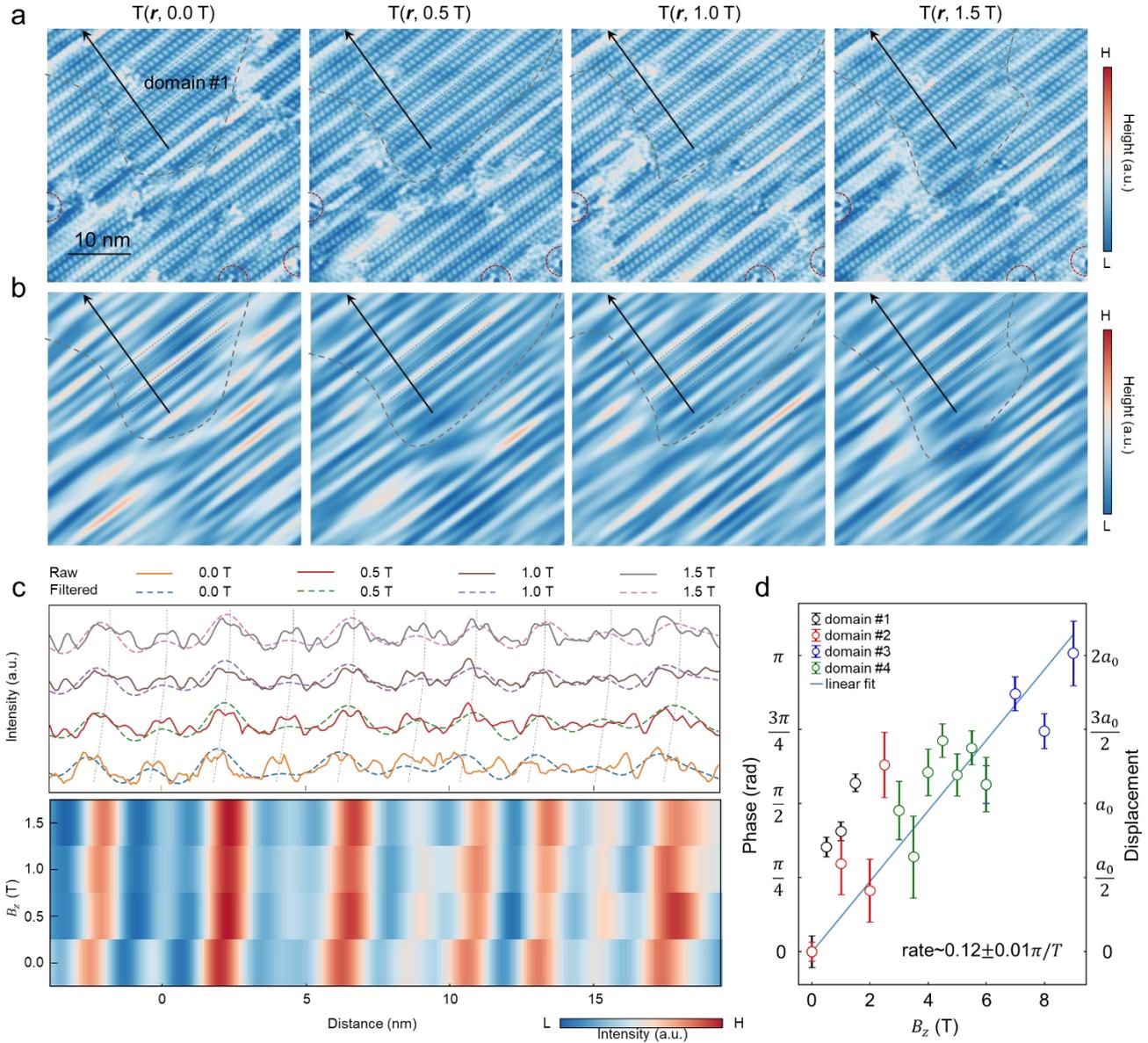

**Fig. 5. Sliding CDW under magnetic field**. **a**, STM topographic images T($r$, $B_z$) of mixed $4a_0 \times \sqrt{3}a_0$ ad $8a_0 \times \sqrt{3}a_0$ CDW region under $B_z$=0, 0.5, 1, and 1.5 T, showing continuous evolution within a single domain (domain #1). Measurement conditions: $V_b$=-50 mV, $I$=1 nA. Dotted lines mark stripe maxima at 0 T for reference. **b**, Corresponding filtered topographic image T$_f$ ($r$, $B_z$) obtained by isolating $q_{4a_0}$ and $q_{8a_0}$ components and performing inversed FT, from the same regions as in (a). Dotted lines again mark stripe maxima at 0 T. **c**, Top: line profiles extracted from (a) along the black arrows, showing stripe shifts. Bottom: intensity maps from line profiles of (b), highlighting the displacement of CDW crests. **d**, Magnetic-field-dependent stripe phase evolution (stripe displacement) extracted by fitting with $\cos(q_{4a_0} \cdot r + \varphi)$. Data are combined from four domains exhibiting continuous evolution, assuming a common phase under overlapping field conditions.